\documentclass{eas}
\usepackage{graphicx}
\usepackage{natbib}
\usepackage{amsmath}
\usepackage{amssymb}
\usepackage{pstricks}
\renewcommand{\vec}{\boldsymbol}
\newcommand{\rnnut}{\ensuremath{\vec{r},\vec{n},\nu,t}}
\newcommand{\mn}{\ensuremath{{\mu\nu}}}
\newcommand{\deriv}[2]{\frac{{\mathrm d}#1}{{\mathrm d}#2}}
\newcommand{\derivl}[2]{{{\mathrm d}#1}/{{\mathrm d}#2}}
\newcommand{\pderiv}[2]{\frac{{\partial}#1}{{\partial}#2}}
\newcommand{\pderivl}[2]{{{\partial}#1}/{{\partial}#2}}

\newcommand{\zav}[1]{\left(#1\right)}
\newcommand{\czav}[1]{\left.#1\right|}
\newcommand{\hzav}[1]{\left[#1\right]}
\newcommand{\szav}[1]{\left\{#1\right\}}

\newcommand{\pul}{\ensuremath{\frac{1}{2}}}
\newcommand{\der}{\ensuremath{\,\mathrm{d}}}
\newcommand{\dmu}{\ensuremath{\,\mathrm{d}\mu}}
\newcommand{\dnu}{\ensuremath{\,\mathrm{d}\nu}}

\newcommand{\NL}{\ensuremath{\mathrm{NL}}}
\newcommand{\ND}{\ensuremath{\mathrm{ND}}}
\newcommand{\NF}{\ensuremath{\mathrm{NF}}}

\newcommand{\dtau}{\ensuremath{\,\mathrm{d}\tau}}

\defcitealias{nltijk}{these proceedings}
\defcitealias{DETAILcode}{these proceedings}
\defcitealias{Kielcode}{these proceedings}
\defcitealias{nicelm}{these proceedings}

\TitreGlobal{Non-LTE Line Formation for Trace Elements in Stellar
Atmospheres}
\begin{document}

\title{Radiative transfer in stellar atmospheres}
%
\author{Ji\v{r}\'{\i} Kub\'at}\address{Astronomick\'y \'ustav AV \v{C}R,
251 65 Ond\v{r}ejov, Czech Republic}
\begin{abstract}
This review presents basic equations for the solution of the NLTE
radiative transfer problem for trace elements and methods for its
solution are summarized.
The importance of frequency coupling in radiative transfer in stellar
atmospheres is emphasized.
\end{abstract}
\maketitle

\section{Introduction}

A stellar atmosphere is the transition between the dense optically thick
stellar core and the transparent interstellar medium.
It is the boundary layer of a star and it is the only part of the star
we can see directly.
Knowledge of the stellar atmosphere is a gate to the investigation of
the stellar interior, and, consequently, to stellar evolution processes.
The radiation passing through it serves not only as an information tool
describing what is happening there, but it also significantly modifies
the stellar atmosphere itself.
This makes the task of determining the emergent radiation from
the star highly nonlinear and therefore also complicated.

Since the stellar atmosphere is a transition from the optically thick
stellar interior to optically thin interstellar medium, the
corresponding radiative transfer equation for the whole atmosphere
cannot be subject to global simplifications.
For the optically thick part of the atmosphere the diffusion
approximation is valid.
Radiation propagates forward slowly resembling the process of diffusion.
This approximation greatly simplifies the radiation transfer equation.
However, the diffusion approximation is unusable for the optically thin
part of the atmosphere.
The situation there is again simple from another point of view.
Since the medium is optically thin, very little interaction with matter
occurs there, which often allows us to neglect absorption and to take
into account only emission.
Between these two relatively simple regions there is a transitional
region, where neither diffusion nor optically thin approximations are
valid.
We have to solve the full set of radiative transfer equations together
with many constraint equations, which introduce the dependence of the
absorption and emission coefficients on radiation and which make the
problem to be solved highly nonlinear.

Radiation carries information about the medium where it was created.
In stellar atmospheres, it has an additional role.
Since the photon mean free path is much larger than the particle mean
free path, radiation in stellar atmospheres influences the matter far
from the place where it is formed, sometimes very significantly.
It changes the population numbers of particular atomic energy levels,
sometimes very far from the equilibrium values.
These changes then influence spectral lines, which are observed by us.
Radiation also transfers energy causing heating (or cooling) of specific
parts of the stellar atmosphere.
In some cases radiation has such an enormous effect on the atmospheric
matter that it becomes the basic reason for a stellar wind, which is
then called as 'radiatively driven'.
All these crucial effects of radiation influence the transfer of
radiation, which becomes strongly nonlinear.

\section{Model stellar atmosphere}

Construction of a stellar atmosphere model is a standard task
of stellar atmosphere physics.
It may be considered as a task to determine space distribution of basic
macroscopic physical quantities ($\vec{r}$ is the position vector and
$\nu$ is the frequency), temperature $T(\vec{r})$, electron number
density $n_e(\vec{r})$, density $\rho(\vec{r})$, velocity
$\vec{v}(\vec{r})$, radiation field $J(\nu,\vec{r})$, population numbers
$n_i(\vec{r})$ of the level $i$, and others by solving equations for
energy equilibrium (which determines $T$), radiative transfer ($J$),
statistical equilibrium ($n_i$), state equation ($n_e$), continuity
equation ($\rho$), equation of motion ($\vec{v}$), and possibly also
some other.
For the simpler case of a static atmosphere in radiative equilibrium,
equations of continuity and
motion are replaced by the hydrostatic equation (which determines
$\rho$) and the energy equation simplifies to the equation of radiative
equilibrium.
Even in the static case the system of equations is quite complicated 
and further simplifications are often used.
Once the atmospheric structure is known, detailed radiation field
specific intensity $I(\nu,\vec{n})$ for each direction $\vec{n}$ can be
calculated by a simple formal solution of the radiative transfer
equation.

The final goal of stellar atmosphere modelling is straightforward.  We
want to compare the theoretical emergent radiation with that detected by
observational instruments.
In principle, it is possible to determine a detailed emergent radiation
energy distribution from scratch (together with a model atmosphere).
Due to the need to know many details of the emergent spectrum, both the
number of variables which need to be determined and consequently the
number of equations to be solved become enormous.
To make the problem tractable, this task is usually not solved at once,
but it is divided into particular steps.
First, the model atmosphere (preferrably NLTE) is calculated by taking
into account all \emph{substantial} contribution of both physical
processes and atomic data.
Minor chemical elements and weak lines need not be taken into account in
a great detail providing they do not influence the atmosphere structure
significantly.
Then the emergent radiation is calculated for a \emph{given} model
atmospheric structure taking into account all details of the chemical
composition and line profile structure including the weakest lines.

Alternatively, it is possible to insert an intermediate step after the
model atmosphere calculation, namely we may determine the NLTE level
populations of some ions not significantly influencing the
atmospheric structure (trace elements) more accurately, and to solve
the coupled set of radiative transfer and statistical equilibrium
equations for them.
For this restricted case with a given model atmosphere (i.e.  $T$ and
$\rho$ are fixed) we do not solve the radiative and hydrostatic
equilibrium equations and we solve only the radiative transfer and
statistical equilibrium equations.
This is the case of the NLTE line formation of trace elements in stellar
atmospheres, which is the subject of these proceedings.
In this case, we have to account for full angle and frequency dependence
of radiation.
In the subsequent step of emergent radiation calculation, they are
considered as given.
Note that in all steps the radiative transfer equation has to be solved.

\section{Basic equations}
\subsection{Radiative transfer equation}

The full radiative transfer equation
for the specific intensity of radiation $I(\rnnut)$
in three dimensions reads \citep[][Eq.\,2-24]{mih}
\begin{equation}\label{rte3d}
\frac{1}{c}\pderiv{I(\rnnut)}{t} + (\vec{n} \cdot \nabla)
I(\rnnut) = \eta(\rnnut) - \chi(\rnnut) I(\rnnut).
\end{equation}
Here $\eta(\rnnut)$ is the emissivity and $\chi(\rnnut)$ is the
absorption coefficient (opacity).
In some mathematicians' language this equation becomes a ``7D''
equation (3 spatial dimensions, 2 directional cosines, time, and
frequency).
However, in the following, we shall use the notation ``xD'' referring
only to the spatial dimensions.
To describe fully the stellar atmosphere with all its properties and
possible features (imagine a picture of the closest star -- the Sun),
it is necessary to use the full 3D radiative transfer equation.
This is a huge task and it may hardly give any reasonable results, even
using contemporary fast computers.

However, in some situations it is possible to use various simplifying
assumptions.
The basic common assumption is the assumption of stationarity, which
means that all $\pderivl{}{t}\rightarrow 0$, thus, all quantities are
time independent.
This assumption is used very often, however, it does not mean that
it is necessarily always valid.
Observational data are always averages over a certain time interval,
which is usually much longer than the typical time scale for rapid
changes in the atmosphere.
This is, of course, true for distant objects, like most stars (where we
have to collect enough light),
but it fails for the closest star, our Sun.
Nonetheless, in the following, we shall assume that the stationarity
assumption is justified.

\begin{figure}[h]
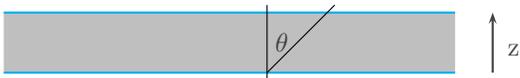

\pspicture(0mm,0mm)(10mm,10mm)
\psframe*[linecolor=lightgray](0mm,0mm)(60mm,8mm)
\psline[linecolor=cyan](0mm,0mm)(60mm,0mm)
\psline[linecolor=cyan](0mm,8mm)(60mm,8mm)
\psline[linecolor=darkgray,arrows=->](65mm,0mm)(65mm,8mm)
\rput[linecolor=cyan](67mm,3mm){\darkgray z}
\psline[linewidth=0.15mm](35mm,-1mm)(35mm,9mm)
\psline[linewidth=0.15mm](35mm,0mm)(44mm,9mm)
\rput(37mm,4mm){\darkgray $\theta$}
\endpspicture
\caption{Schematic picture of the plane-parallel stellar atmosphere.
Specific intensity of radiation depends on a $z$-coordinate, on
the angle cosine $\mu=\cos\theta$, and on frequency $\nu$.}
\label{jk1refschema}
\end{figure}

If the stellar atmosphere is very thin compared to the stellar radius,
then its curvature may be neglected and close parallel rays stay almost
nearly parallel throughout the atmosphere.
Then we can assume that the atmosphere is an infinite thin horizontally
homogeneous plane.
Such model stellar atmosphere is called \emph{plane-parallel}
(Fig.\,\ref{jk1refschema}).
This is a very common assumption, which simplifies the geometry
of a problem, it allows to use a 1D radiative transfer equation instead
of the full 3D one.
Then, the radiative transfer equation takes the form (in the following
we will often use the common notation convention
$I(\nu)\rightarrow I_\nu$)
\begin{equation}\label{rte1dpp}
\mu \deriv{I_\mn(z)}{z}
= \eta_\nu(z) - \chi_\nu(z) I_\mn(z)
= \chi_\nu \hzav{S_\nu(z) - I_\mn(z)},
\end{equation}
where $\mu=\cos\theta$ is the angle cosine of the ray and
$S_\nu=\eta_\nu/\chi_\nu$ is the source function.
Note that the one-dimensionality of
equation \eqref{rte1dpp} means
that the physical quantities depend only on one coordinate ($z$).
The radiation field fills the whole space in and full angle dependence
is taken into account, of course, with inherent symmetries, which allow
to use only one angle cosine.

If the stellar atmosphere is not thin compared to the stellar radius,
the plane-parallel assumption may become invalid (e.g. for limb
radiation calculations).
Then another very common simplifying assumption, which also leads to
one-dimensional atmosphere, may be used.
It is the assumption of spherical symmetry.
This assumption is useful for the treatment of extended atmospheres.
The radiative transfer equation \eqref{rte3d} then takes the form
\begin{equation}\label{rte1ds}
\mu \pderiv{I_\mn(r)}{r} + \frac{1-\mu^2}{r} \pderiv{I_\mn(r)}{\mu}
= \eta_\nu(r) - \chi_\nu(r) I_\mn(r).
\end{equation}
Here $r$ is the radial coordinate and $\mu=\cos\theta$ is again the
angle cosine of the ray, as in the plane-parallel case.

Alternatively, 
if we introduce the so-called variable Eddington factor
$f_\nu= \pul \int_{-1}^1 \mu^2 I_\mn \dmu/J_\nu$ \citep{vef},
we may write this equation as a 2nd-order differential equation for the
first moment of the intensity
$J_\nu=\pul \int_{-1}^1 I_\mn \dmu$ (the mean intensity),
\begin{equation}\label{rtm1dpp}
\deriv{^2 \hzav{f_\nu(z) J_\nu(z)}}{\tau_\nu^2}
= J_\nu(z) - S_\nu(z),
\end{equation}
where $\dtau_\nu = -\chi_\nu \der z$ is the optical depth.
Opacity ($\chi_\nu$) and emissivity ($\eta_\nu$) in the radiative
transfer equation depend not only on local temperature and density, but
also on radiation, which may propagate much longer distances than the
mean free path of particles.
This makes the problem non-local and non-linear.

Equation \eqref{rtm1dpp} has to be supplemented by boundary
conditions, which for the case of the stellar atmosphere may be
\begin{subequations}\label{okraje}
\begin{align}
&\deriv{\hzav{f_\nu(z) J_\nu(z)}}{\tau_\nu} = g_\nu(z) - H_\nu^-
& \text{for the upper boundary and}
\\
&\deriv{\hzav{f_\nu(z) J_\nu(z)}}{\tau_\nu} = H_\nu^+ + g_\nu(z)
& \text{for the lower boundary,}
\end{align}
\end{subequations}
where $g_\nu = {\int_0^1 \mu \zav{I_\mn^+ - I_\mn^-} \dmu}/{J_\nu}$,
and $H_\nu^+$, $H_\nu^-$ are incident fluxes at corresponding
boundaries.
It is common to assume zero incident radiation ($H_\nu^-=0$) and
diffusion approximation at the lower boundary
($H_\nu^+=[1/3][\derivl{B_\nu}{\tau_\nu}]$).

Opacity and emissivity (and then the source function $S_\nu$) in stellar
atmospheres can be calculated using the expressions
\citep[see][Eqs.\,7-1 and 7-2]{mih}
\begin{multline}\label{opacita}
\chi_\nu =
\sum_i \sum_{l>i} \hzav{n_i - \frac{g_i}{g_l} n_l} \alpha_{il}(\nu) +
\sum_i \zav{n_i - n_i^\ast e^{-\frac{h\nu}{kT}}} \alpha_{ik}(\nu) + \\
\sum_k n_e n_k \alpha_{kk} (\nu,T) \zav{1-e^{-\frac{h\nu}{kT}}} +
n_e \sigma_e
\end{multline}
\begin{multline}\label{emisivita}
\eta_\nu = \frac{2h\nu^3}{c^2} \hzav{
\sum_i \sum_{l>i} n_l \frac{g_i}{g_l} \alpha_{il}(\nu) +
\sum_i n_i^\ast \alpha_{ik}(\nu) e^{-\frac{h\nu}{kT}} +
\right. \\ \left.
\sum_k n_e n_k \alpha_{kk} (\nu,T) e^{-\frac{h\nu}{kT}}}.
\end{multline}
The first term on the right hand sides of both equations \eqref{opacita}
and \eqref{emisivita} corresponds to line opacity
(bound-bound transitions), the second one to bound-free transitions,
the third one to free-free transitions, and the last one (only in
\ref{opacita}) to electron scattering.
In these equations, $n_i$, $n_l$, and $n_k$  are level populations
($n_i^\ast$ is the equilibrium value defined with respect to the ground
level of the next higher ion), $g_i$ and $g_l$ are statistical
weights, $\alpha$ denotes the cross section for the corresponding
transition, $n_e$ is the electron number density, and $\sigma_e$ is
the Thomson scattering cross section.
Note that all level population may be heavily influenced by the
radiation field.

For the case where we can use the equilibrium values of population
numbers (the so-called `local thermodynamic equilibrium -- LTE'),
quantities $n_i$ can be replaced by their equilibrium values $n_i^\ast$,
which depend only on local values of temperature $T$ and electron
density $n_e$ through the Saha ionization and Boltzmann excitation
equilibrium laws.
This greatly simplifies the problem, because both opacity and emissivity
then depend only on local temperature and density, and are independent
of radiation.
Then the radiative transfer equation \eqref{rtm1dpp} is linear in
$J_\nu$.

\subsection{Equations of statistical equilibrium}

However, such a simplified description is not generally valid.
In stellar atmospheres, the conditions may be far from thermodynamic
equilibrium and Saha-Boltzmann equations for level population numbers
cannot be used.
In this case, level populations are determined using the equations of
statistical equilibrium, 
\begin{equation}\label{eseil}
\sum_{l\ne i} \hzav{n_l \zav{R_{li} + C_{li}}
                  - n_i \zav{R_{il} + C_{il}}} = 0,
\quad\quad i=1,\dots \NL
\end{equation}
This is the set of {\NL} equations for each explicitly considered level.
Since these equations are linearly dependent, an additional equation has
to be used.
Usually the charge or particle conservation equations are used.
In the case of trace elements the equation which defines the abundance
of the element with respect to some reference element (usually
hydrogen, but not necessarily) may be used.

The quantities $R_{il}$ and $C_{il}$ in the equation \eqref{eseil} are
radiative and collisional rates, respectively, for transitions between
energy levels $i$ and $l$.
Collisional rates,
\begin{equation}\label{colrates}
n_i C_{il}
= n_i n_e q_{il}(T),
\end{equation}
where $q_{il}$ is a function of temperature only, do not depend on the
radiation field (note that $n_l^\ast C_{li} = n_i^\ast C_{il}$).
On the other hand, both upward ($R_{il}$) and downward ($R_{li}$)
radiative rates,
\begin{subequations}\label{radrates}
\begin{align}
n_i R_{il} & = n_i 4\pi \int \frac{\alpha_{il}(\nu)}{h\nu} J_\nu \dnu
\\
n_l R_{li} &
= n_l \frac{g_i}{g_l} 4\pi \int \frac{\alpha_{il}(\nu)}{h\nu}
\zav{\frac{2h\nu^3}{c^2} + J_\nu}
\dnu,
\end{align}
\end{subequations}
depend explicitly on the mean radiation intensity $J_\nu$.
Detailed expressions for radiative rates can be found in the next chapter
\citep[\citetalias{nltijk}]{nltijk}.

\subsubsection{Two-level atom}

The basic principles are usually best shown using very simple, so called
textbook cases.
The concept of a two-level atom falls within them.
Let us assume an atom having only two levels \citep[see][]{mih}.
In this case the equations of statistical equilibrium \eqref{eseil}
reduce to
\begin{equation}\label{ese2l}
n_2 \zav{R_{21} + C_{21}} - n_1 \zav{R_{12} + C_{12}} = 0
\end{equation}
For this case it may be derived that the source function
\begin{equation}\label{vydatnost2l}
S_\nu = \zav{1-\varepsilon} \int \varphi_\nu J_\nu \dnu + \varepsilon
B_\nu,
\end{equation}
where $\varphi_\nu$ is the line profile function, $B_\nu$ is the Planck
function, and $\varepsilon=\varepsilon^\prime/(1+\varepsilon^\prime)$,
and $\varepsilon^\prime=C_{21} \hzav{1-\exp\zav{-{h\nu}/{kT}}}/A_{21}$.
Equation \eqref{vydatnost2l} clearly shows that the physical processes
can be separated into the scattering ones, which depend on the radiation
field $J_\nu$, and to the thermal ones, which depend on the local Planck
function.
The lower the value of $\varepsilon$, the less the LTE assumption is
acceptable.

\subsection{Final radiative transfer equation with a constraint of
statistical equilibrium}

If we explicitly emphasize the quantity dependences in
the equation \eqref{rte1dpp}, it then takes the form
\begin{equation}\label{rtm1dppexp}
\deriv{^2 \hzav{f_\nu(z) J_\nu(z)}}{\tau_\nu^2}
= J_\nu(z) - S_\nu(z,J_{\nu^\prime})
\end{equation}
which shows its nonlinearity in the specific radiation intensity $J$.
Using $\nu^\prime$ instead of $\nu$ indicates that the radiative
transfer in a particular frequency generally depends on the radiation
field at all other frequencies.

Equation \eqref{rtm1dppexp} with the boundary conditions \eqref{okraje}
has to be solved using opacity \eqref{opacita} and emissivity
\eqref{emisivita}, together with the statistical equilibrium equations
\eqref{eseil} using both radiative \eqref{radrates} and collisional
\eqref{colrates} rates.
These equations together describe the basic NLTE radiative transfer
problem which has to be solved in stellar atmospheres.

\section{Frequency coupling of radiation}

Let us turn our attention to another aspect of the radiative transfer
problem, namely radiation coupling.
One of the most striking features of radiation is the coupling of
distant places, which would otherwise remain uncoupled.
This coupling may occur over very long distances within stellar
atmospheres.
This long-distance coupling is described by the radiative transfer
equation (\ref{rte3d}, \ref{rte1dpp}, \ref{rte1ds}).
Besides long distance coupling, scattering causes coupling between
different radiation propagation directions.

An important aspect of radiation is the frequency coupling, which is
crucial in stellar atmospheres.
Although we succeeded to reduce the geometry complexity of the problem
significantly, we have to retain full frequency dependence of the
equations \eqref{rte1dpp} and \eqref{rte1ds}, since the radiative
transfer in individual frequencies influences transfer in many other
frequencies.
Each of the equations \eqref{rte1dpp} or \eqref{rte1ds} represents
a set of frequency coupled equations.

The trivial case is that of no interaction of radiation with matter.
Then the specific intensity $I$ remains constant and there is no
frequency coupling.
The simplest case is that of pure coherent scattering, either continuum
(like electron scattering) or in lines.
In this case, photons may change direction, but not frequency and the
radiative transfer equation has the form
\begin{equation}
\mu \deriv{I_\mu}{\tau} = I_\mu - \pul \int_{-1}^1
\sigma_{\mu\mu^\prime} I_{\mu^\prime}\dmu^\prime,
\end{equation}
where $\sigma_{\mu\mu^\prime}$ is the scattering cross-section.
This kind of radiative transfer problem forms a significant part of the
Chandrasekhar's Radiative Transfer book \citep{chanrt}.
In this case the radiative transfer at individual frequencies may be
solved independently of other frequencies.

\subsection{Frequency coupling within spectral lines}

In a spectral line, frequencies of absorbed and emitted photon may
differ within the line profile, which for the case of pure natural
broadening (Lorentz profile) is
\begin{equation}\label{prflor}
\varphi_\nu = \frac{\dfrac{\Gamma}{4\pi^2}}{\zav{\nu-\nu_0}^2 +
\zav{\dfrac{\Gamma}{4\pi}}^2},
\end{equation}
where $\nu_0$ is the central line frequency and $\Gamma$ is the
damping parameter.
For the case of complete or partial frequency redistribution
there is a coupling between them via the redistribution process.

Thanks to thermal motions the spectral lines are subject to Doppler
broadening.
As a consequence, even in the case of coherent scattering (which takes
place in the atomic
frame), thermal motions may cause the difference between absorbing and
emitting frequency.
The broadened line profile for the case of atomic frame coherent
scattering (Doppler profile) is
\begin{equation}
\varphi_\nu = \frac{1}{\Delta\nu_D \sqrt{\pi}}
\exp \zav{\frac{\nu-\nu_0}{\Delta\nu_D}}
\end{equation}
where the Doppler half-width
$\Delta\nu_D = \zav{\nu_0/c} \sqrt{{2kT}/{m}}$.
For the case of a naturally broadened line
\eqref{prflor} the profile is (Voigt
profile)
\begin{equation}
\varphi_\nu = \frac{1}{\Delta\nu_D \sqrt{\pi}} H \zav{a,v},
\end{equation}
where $H \zav{a,v}$ is the Voigt function \citep[see][Eq.\,9-34]{mih}.
Thermal broadening is dominant in stellar atmospheres, which strongly
couples frequencies within each line.
In addition other additional broadening mechanisms, line Stark or
collisional broadening, may introduce further frequency coupling.

To further illustrate the effect of lines on frequency coupling, we
use the example of a two-level atom with partial frequency
redistribution.
The radiative transfer equation for this case is
\begin{equation}\label{jk1dvouhl}
\mu \deriv{I_\nu}{z} = \frac{h\nu}{4\pi}
\Bigl[{-n_l \varphi_\nu B_{lu} I_\nu +
n_u \psi_\nu \zav{A_{ul}+B_{ul}I_\nu}}\Bigl],
\end{equation}
where the emission profile
$\psi(\nu) = \int r\zav{\nu^\prime,\nu} \varphi(\nu^\prime)
\dnu^\prime$, $r\zav{\nu^\prime,\nu}$ is the redistribution function,
$n_l$ and $n_u$ are the occupation numbers of the lower ($l$) and upper
($u$) levels, and $A_{ul}$, $B_{ul}$ and $B_{lu}$ are the Einstein
coefficients.
All line frequencies, which are coupled via
Equation \eqref{jk1dvouhl}, have to be solved together.

\subsection{Frequency coupling across lines}

A photon after being absorbed, may be then re-emitted in the same line
(this option was already mentioned in the preceding Section), or it may
be emitted in a different line (or continuum transition), or it may be
destroyed by collisional transition.
In the latter case the photon's energy is converted to heat.
As a reverse process, 
photons may be emitted after collisional excitation, which causes
cooling.

Each absorption or emission changes the distribution of atomic
excitation states, which causes the change in the opacity.
The changes are reflected by the set of statistical equilibrium
equations \eqref{eseil}.
However, the radiative rates \eqref{radrates} depend on radiation field.
Consequently, we have to solve the equations 
\eqref{rte1dpp} and \eqref{eseil} simultaneously.

This coupling, which causes absorption of radiation in one part of the
spectrum and its subsequent emission in a different spectral part, peaks
in the effect usually referred to as line-blanketing.
Spectral lines are not distributed across the frequency spectrum evenly.
There appears a huge amount of lines in the UV spectral region.
In addition, the radiative flux in this region is large.
As a consequence, the radiation is absorbed in UV and re-emitted in the
visual and infrared regions, where the opacity is lower.
It is caused mostly by metallic lines dominated by Fe and Ni.
The effect of line blanketing hes been studied by a number of authors
\citep[e.g.][and references
therein]{kurmod,karbous,nltekiel,himi,nlteblhot1,nlteblo,nlteblb}.

\subsection{Frequency coupling in moving atmospheres}

For expanding atmospheres ($\vec{v}(\vec{r}) \ne 0$, $\derivl{v}{r}>0$)
another type of frequency coupling occurs.
Since the process of atomic absorption and emission (scattering) takes
place in a rest frame of an atom (i.e. in a co-moving frame), when it is
observed from the reference frame connected with a star (the observer's
frame), Doppler shift changes the frequency of lines.
The radiative transfer equation in the observer frame has the form
\eqref{rte1dpp}, but the opacity ($\chi_\mn(z)$) and emissivity
($\eta_\mn(z)$) coefficients are angle dependent.
The angle dependence is caused by the Doppler shift
$\nu^\prime = \nu \zav{1 - {\vec{n}\cdot\vec{v}}/{c}}
= \nu \zav{1-\mu {v}/{c}}$.
Equivalently, we may write the radiative transfer equation in the
co-moving frame (we change the notation $\mu^\prime \rightarrow \mu$,
$\nu^\prime \rightarrow \nu$)
\begin{displaymath}
\mu \pderiv{I_\mn(z)}{z}
- \hzav{\frac{\mu^2 \nu}{c} \pderiv{v}{z}} \pderiv{I_\mn(z)}{\nu}
= \eta_\nu(z) - \chi_\nu(z) I_\mn(z).
\end{displaymath}
The second term on the left hand side clearly shows the frequency
coupling caused by a derivative $\pderivl{I_\mn(z)}{\nu}$.
If we can neglect this term, we obtain the static radiative transfer
equation.
On the other hand, if we can neglect the first term, we obtain the
so-called Sobolev approximation \citep[see][]{cashydr}.

\section{Discretization}

Since the solution of the combined equations of radiative transfer and
statistical equilibrium is performed numerically, we have to switch from
continuous independent variables to a discrete set of representative
points.
Instead of the depth variable $z$ we use depth points indexed by $d$.
Since the density in stellar atmospheres varies by several orders of
magnitude, the advantageous choice is to space the depth points
equidistantly in $\log m$ ($m$ is the column mass depth, $\der m= -
\rho \der z$) or $\log \tau$,
where $\tau$ is some suitable optical depth, e.g. Rosseland mean optical
depth or an optical depth at some representative wavelength.
The discretization of frequency $\nu \rightarrow n$ has to be done to
resolve all spectral lines and continuum edges, consequently, it
cannot be equidistant.
On the other hand, for plane-parallel atmospheres it is advantageous to
use Gaussian quadrature for angle integration, which then defines the
rays along which the radiative transfer equation is solved.

\paragraph{Discretized radiative transfer equation}
Discretizing the radiative transfer equation \eqref{rtm1dpp} for the
frequency point $n$ we obtain
\cite[see also][]{feautrier}
\begin{subequations}
\begin{align}
&a_d f_{d-1} J_{d-1} + \zav{b_d f_d + 1} J_d + c_d f_{d+1} J_{d+1} = S_d
\quad\text{for $d=2,\dots,\ND-1$,}
\\
&\zav{b_1 f_1 + g_1 + 1} J_1 + c_1 f_2 J_2 = H^- + S_1,
\quad\text{and}
\\
&a_\ND f_{\ND-1} J_{\ND-1} + \zav{b_\ND f_\ND + g_\ND + 1} J_\ND = H^+ +
S_\ND.
\end{align}
\end{subequations}
In these equations,
\begin{subequations}
\begin{align}
a_d &= - \hzav{\pul \zav{\Delta \tau_{d-\pul} +
\Delta \tau_{d+\pul}} \Delta \tau_{d-\pul} }^{-1}
\\
c_d &= - \hzav{\pul \zav{\Delta \tau_{d-\pul} +
\Delta \tau_{d+\pul}} \Delta \tau_{d+\pul}}^{-1}
\\
b_d &= - a_d - c_d,
\\
c_1 &= - \zav{\Delta \tau_\frac{3}{2}}^{-1},
\quad
b_1=-c_1,
\\
a_\ND&= - \zav{\Delta \tau_{\ND-\pul}}^{-1},
\quad\text{and\quad}
b_\ND = -a_\ND,
\end{align}
\end{subequations}
where $\Delta \tau_{d-\pul}= \tau_d - \tau_{d-1}$.
The above equations refer to the case of 2nd-order differences, which is
the most stable way of discretizing the differential radiative transfer
equation.
We may also use either splines or Hermite differences
\cite[see][]{hermite}, which are more
accurate, but sometimes we run into troubles since they are less stable
and oscillatory behaviour occasionally occurs.

\paragraph{Discretized frequency integration}
The integral both across the line profile and over continuum frequencies
is replaced by a quadrature sum
\begin{equation}
\int J(\nu) \dnu \rightarrow \sum_{n=1}^\NF w_n J_n.
\end{equation}
For lines, it is necessary to ensure that
$\int \phi(\nu) \dnu = \sum_{n=1}^\NF w_n \phi_n = 1$.
If this condition is not fulfilled, it is obligatory to renormalize
$w_n$.
Otherwise, an artificial unphysical source or sink of photons appears,
which may dramatically change results.

To ensure correct treatment of ionization edges, it is advantageous to
use two close frequency points which differ only by machine accuracy,
one before and one after the ionization edge (Rauch, private
communication).
This ensures exact treatment of all contributions to the particular
ionization rate and also to other rates where the continuum edge is a
standard continuum frequency point.

\paragraph{Discretized angle integration}
The integral over the angles in the calculation of the intensity moments
is also replace by quadrature sums,
\begin{equation}
\int I(\mu) \dmu \rightarrow \sum_{m=1}^\NF w_m I_m.
\end{equation}
Usually, 3 points are sufficient if Gaussian angle quadrature
\citep[see][Section\,22]{chanrt} is used.
Also in this case the check of correct normalisation is crucial.

\paragraph{Discretized equations of statistical equilibrium}
For all $d=1,\dots,\ND$, $i=1,\dots,\NL$ we may write
\begin{equation}
\sum_{l\ne i} \szav{(n_l)_d \hzav{
(R_{li})_d + (C_{li})_d} - (n_i)_d \hzav{
(R_{il})_d
+ (C_{il})_d}} = 0,
\end{equation}
where the radiative rates are given by
\begin{subequations}
\begin{align}
&(n_i)_d (R_{il})_d = (n_i)_d 4\pi \sum_{n=1}^\NF w_n
\frac{\zav{\alpha_{il}}_n}{h\nu_n} J_{dn},
\\
&(n_l)_d (R_{li})_d
= (n_l)_d \frac{g_i}{g_l} 4\pi \sum_{n=1}^\NF w_n
\frac{\zav{\alpha_{il}}_n}{h\nu_n}
\zav{\frac{2h\nu_n^3}{c^2} + J_{dn}}.
\end{align}
\end{subequations}

\section{Formal solution of the radiative transfer equation}

Formal solution of the RTE is the next step after calculation of the
model atmosphere and the occupation numbers.
It is any radiative transfer solution for \emph{given} $\chi_\nu$ and
$\eta_\nu$ (given $S_\nu$) and it is relatively relatively simple.
In this step the equation \eqref{rte1dpp} is directly solved, or it is
sometimes rewritten to the 2nd-order form.
The latter possibility has become more popular in one-dimensional model
atmospheres.
The formal solution is crucial for the total accuracy of the whole
problem solution.

\subsection{Feautrier solution}

The most widely used radiative transfer equation solution for static
plane-parallel atmospheres is the solution of the second-order system
introduced by \cite{feautrier}.
In this scheme we combine opposite directions along a ray represented by
specific intensities $I^+$ and $I^-$.
Introducing ``Feautrier variables'', i.e. their symmetric and
anti-symmetric means,
\begin{subequations}
\begin{align}
u_\mn &= \pul \zav{I^+_\mn+I^-_\mn}
\intertext{and}
v_\mn &= \pul \zav{I^-_\mn+I^-_\mn},
\end{align}
\end{subequations}
we obtain the transfer equation
\begin{equation}
\deriv{^2 u_\mn}{\tau_\mn^2} = u_\mn - S_\nu
\end{equation}
supplemented by appropriate boundary conditions.

\subsection{Short characteristics}

\begin{figure}[b]
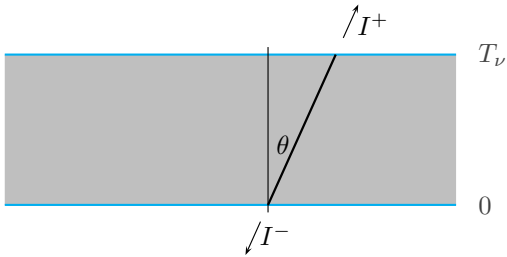

\pspicture(0mm,-3mm)(25mm,22mm)
\psframe*[linecolor=lightgray](0mm,0mm)(60mm,20mm)
\psline[linecolor=cyan](0mm,0mm)(60mm,0mm)
\psline[linecolor=cyan](0mm,20mm)(60mm,20mm)
\rput[linecolor=cyan](63mm,0mm){\darkgray 0}
\rput[linecolor=cyan](63mm,20mm){\darkgray $T_\nu$}
\psline[linewidth=0.15mm](35mm,-1mm)(35mm,21mm)
\psline[linewidth=0.3mm](35mm,0mm)(44mm,20mm)
\rput(37mm,8mm){$\theta$}
\psline[linewidth=0.15mm,arrows=->](45mm,22.22mm)(47mm,26.67mm)
\psline[linewidth=0.15mm,arrows=->](34mm,-2.22mm)(32mm,-6.67mm)
\rput(36mm,-4mm){$I^-$}
\rput(49mm,24mm){$I^+$}
\endpspicture
\caption{Schematic representation of the short characteristics
solution.}
\label{jk1slab}
\end{figure}

The short characteristics method is a first-order method.
If we solve the radiative transfer equation for a finite slab (see
Fig.\,\ref{jk1slab}) with an optical thickness $T_\nu$ along a ray with
an angle $\theta$ ($\mu=\cos\theta$), then by integration of Equation
\eqref{rte1dpp} between 0 and $T_\nu$ we obtain for the outward
direction
\begin{subequations}
\begin{equation}
I_\mn^+ \zav{\tau_\nu} =
I_\mn^+ \zav{T_\nu} e^{-\zav{T_\nu - \tau_\nu}/{\mu}} + \frac{1}{\mu}
\int_{\tau_\nu}^{T_\nu} S\zav{t} e^{-\zav{t - \tau_\nu}/{\mu}} \der t,
\end{equation}
and for the inward direction
\begin{equation}
I_\mn^- \zav{\tau_\nu} =
I_\mn^- \zav{0} e^{-{\tau_\nu}/{\mu}} + \frac{1}{\zav{-\mu}}
\int_0^{\tau_\nu} S\zav{t} e^{-\zav{\tau_\nu - t}/{\mu}} \der t.
\end{equation}
\end{subequations}
The stellar atmosphere is then divided into shells (i.e. finite slabs)
between particular depth points and solved by parts throughout the
atmosphere.

While not being used for the solution of the one-dimensional radiative
transfer equation as much as the Feautrier solution, the power of the
short characteristics method emerges for multidimensional radiative
transfer \citep[see][]{kunauer}.

\section{$\Lambda$~iteration}

We may formally write the formal solution of the equation
\eqref{rtm1dpp}, which was described in the preceding section, as an
operation of an operator $\Lambda$ on the source function
\citep[see][]{mih},
\begin{equation}\label{lameq}
J_\nu = \Lambda_\nu S_\nu.
\end{equation}
Generally, the source function, $S_\nu$, depends on the radiation field
intensity $J_\nu$, which makes the problem non-linear.
The simplest way how to cope with the non-linearity is an iterative
procedure.
Let us assume some starting values for opacity and emissivity, say the
LTE values based on LTE level populations $n_i^\ast$ (i.e. the source
function $S_\nu=B_\nu$).
The opacities and emissivities can be easily calculated from
\eqref{opacita} and \eqref{emisivita}, respectively, using LTE
population values.
Then instead of the radiative transfer equation \eqref{rtm1dppexp},
equation \eqref{rtm1dpp} can be solved (for given $\eta_\nu$ and
$\chi_\nu$, i.e. for given $S_\nu$), which is in fact the formal
solution mentioned in the preceding section.
Then, the equations of statistical equilibrium \eqref{eseil} are solved
with radiative rates calculated for a given radiation field.
New level populations $n_i$ are determined, which are then used for
calculation of new values of $\chi_\nu$ and $\eta_\nu$, and the process
is iterated.

The above iteration scheme may be written as
\begin{equation}\label{lamit}
J_\nu^{(n)} = \Lambda_\nu S_\nu^{(n-1)},
\end{equation}
where $(n)$ and $(n-1)$ denote the iteration numbers.
The occupation numbers are determined using the equation
\begin{equation}\label{jk2lamitese}
\sum_{l\ne i} \szav{n_l^{(n)} \hzav{ R_{li} (J_\nu^{(n)}) + C_{li}} -
n_i^{(n)} \hzav{ R_{il} (J_\nu^{(n)}) + C_{il}}} = 0.
\end{equation}
The level populations $n_i^{(n)}$ are used to calculate the source
function $S_\nu^{(n)}$ and we follow again with solution of
\eqref{lamit}, now for $J_\nu^{(n+1)}$.
The crucial problem of this iterative process is that, for the case of
stellar atmospheres, it converges extremely slowly, it rather stabilises
far from the true solution.
This is caused by the fact that each iteration represents interaction
over one mean free path.
In the optically thick parts of the atmosphere (with a lot of scattering
events) it means extremely slow propagation of information throughout
the stellar atmosphere.
Illustrative examples can be found in \citet{auermrt}.

Although the lambda iteration is unusable for stellar atmospheres,
for optically thin media, like planetary nebulae or circumstellar
envelopes, it may work \citep[see][]{wc49A}.
However, for stellar atmospheres which are optically thick, different
approaches have to be found.

\section{Complete linearization}\label{linearizace}

The complete linearization method (the Newton-Raphson method) was
introduced to the theory of stellar atmospheres by \cite{amnlte3} for
the case of a full solution of a NLTE model plane-parallel atmosphere,
which means simultaneous solution of equations of radiative transfer,
statistical equilibrium, radiative equilibrium, and hydrostatic
equilibrium in one spatial dimension.
For the simpler task, when the equations of hydrostatic and radiative
equilibrium are not solved and only common solution of the radiative
transfer equation together with the statistical  equilibrium equations
is performed, the complete linearization was described by \cite{clline}.

We have to solve the plane-parallel radiative transfer equation
\eqref{rtm1dpp} together with the set of equations of statistical
equilibrium \eqref{eseil} for radiation intensities $J_\nu$ and level
populations $n_i$.
Since we cannot resolve the radiation field for infinite number of
frequencies, we have to replace the continuous index ($\nu$) by a
discrete one ($i$), $J_\nu \rightarrow J_i$, and solve the radiative
transfer equation only for these frequency points.
The variables $J_i$ and $n_l$ may be formally written as a vector
\begin{displaymath}
\vec{\psi} = \zav{J_1, \dots, J_\NF; n_1, \dots, n_\NL},
\end{displaymath}
which has the dimension $\NF+\NL$.
The equations to be solved may be formally written as
\begin{equation}\label{Fpsi}
\vec{F}\zav{\vec{\psi}} = 0
\end{equation}
If $\vec{\psi}_0$ is the current estimate of the solution, which does
not satisfy the Eq.~\eqref{Fpsi} exactly, then the correct solution can
be obtained by adding a correction
$\delta\vec{\psi}$,
\begin{equation}\label{psirozvoj}
\vec{\psi}=\vec{\psi}_0+\delta\vec{\psi}.
\end{equation}
Inserting it into the equation \eqref{Fpsi} we obtain after expansion
to the first order for the corrections
\begin{equation}
\delta\vec{\psi}
= - \hzav{\pderiv{\vec{F}}{\vec{\psi}}\zav{\vec{\psi}_0}}^{-1} \cdot
{\vec{F}\zav{\vec{\psi}_0}}.
\end{equation}
Then the new values of the vector $\vec{\psi}$ are calculated using
Eq.~\eqref{psirozvoj}.

\subsection{Standard formulation}

After implementing the difference scheme for the Eq.~\eqref{rtm1dppexp},
the linearized radiative transfer equation may be symbolically written
as
\begin{equation}
\mathbb{A}_d \delta \vec{J}_{d-1} + \mathbb{B}_d \delta \vec{J}_d +
\mathbb{C}_d \delta \vec{J}_{d+1} = \vec{L}_d,
\end{equation}
where the vector $\delta\vec{J}_d=\zav{\delta J_{d,1},\dots,\delta
J_{d,\NF}}$ for each $d=1,\dots,\ND$.
$\mathbb{A}_d$, $\mathbb{B}_d$, and $\mathbb{C}_d$ are matrices,
$\mathbb{A}_1=0$, and $\mathbb{C}_\ND=0$.
The right hand side $\vec{L}_d$ contains the source function
$\vec{S}_d$.

The dependence of the source function on the occupation numbers may be
expressed with the help of the derivative
\begin{displaymath}
\delta S_{nd} =
\sum_{l=1}^\NL \czav{\pderiv{S_n}{n_l}}_d \zav{\delta{n}_l}_d.
\end{displaymath}
The combined equation may be then written as
\begin{equation}\label{jk2rslin}
\mathbb{A}_d \delta \vec{J}_{n,d-1} + \mathbb{B}_d \delta \vec{J}_{n,d}
+ \mathbb{C}_d \delta \vec{J}_{n,d+1}
+ \mathbb{D}_d \delta \vec{n}_d
= \vec{M}_{n,d}.
\end{equation}
If we formally write the statistical equilibrium equations \eqref{eseil} as
${\cal A} \cdot \vec{n} = {\vec b}$, where
${\cal A}$ is the rate matrix, $\vec{n}=\zav{n_1,\dots,n_\NL}$, we may
then, for the changes of populations numbers $\delta \vec{n}_d$, write
\begin{equation}
\hzav{\pderiv{{\cal A}}{\vec{n}} \vec{n} - \pderiv{b}{\vec{n}}
+ {\cal A}}_d \delta \vec{n}_d +
\hzav{\pderiv{{\cal A}}{\vec{J}} \vec{n} - \pderiv{b}{\vec{J}}}
\delta \vec{J}_{d} =
{\vec b}_d - {\cal A}_d \cdot \vec{n}_d,
\end{equation}
which may be schematically written as
\begin{equation}\label{jk2eslin}
\mathbb{E}_d \delta \vec{n}_{d} + \mathbb{F}_d \delta \vec{J}_{d} =
\vec{K}_{d}.
\end{equation}
Equations \eqref{jk2rslin} and \eqref{jk2eslin} schematically describe
the problem, which we solve.
Using the linearized equations of statistical equilibrium, we may
express the level population changes as
\begin{equation}
\delta n_{l,d} = \sum_{n=1}^\NF \czav{\pderiv{n_l}{J_n}}_d
\delta (J_n)_d,
\end{equation}
where
\begin{equation}
\pderiv{n_l}{J_n} = \sum_{r=1}^\NL{\cal A}^{-1}_{lr}
\hzav{\pderiv{b_r}{J_n}  -
\sum_{s=1}^\NL \pderiv{{\cal A}_{rs}}{J_n} \cdot n_s}.
\end{equation}
Then the combined radiative transfer $+$ statistical equilibrium
equations may be written as a set of tridiagonal matrices
\begin{equation}
\mathbb{A}^\prime_d \delta \vec{J}_{d-1} + \mathbb{B}^\prime_d \delta \vec{J}_d 
+
\mathbb{C}^\prime_d \delta \vec{J}_{d+1} = \vec{L}^\prime_d,
\end{equation}
where $\delta\vec{J}_d = \zav{\delta J_1,\dots,\delta J_\NF}$,
$d=1,\dots,\ND$ and $\mathbb{A}^\prime_1=0$, $\mathbb{C}^\prime_\ND=0$.

\subsection{Formulation using net radiative brackets}

\cite{clalt} reformulated the application of the complete linearization
method with the help of net radiative brackets.
For each transition $t$ between levels $i$ and $l$, we may introduce the
quantity
\begin{equation}
Z_t = n_i R_{il} - n_l R_{li},
\end{equation}
which describes the net radiative rate.
Linearizing this expression for each depth point $d$ we obtain
\begin{equation}
\zav{\delta Z_t}_d = (n_i)_d (\delta R_{il})_d - (n_l)_d (\delta
R_{li})_d.
\end{equation}
Using $\delta n_i = \sum_t \zav{\pderivl{n_i}{Z_t}} \delta Z_t$ we may
eliminate the population numbers changes $\delta n_i$ and finally
arrive at a system of equations
\begin{equation}
\zav{\vec{1} + {\vec R}_t \pderiv{n_i}{Z_t} + {\vec S}_t
\pderiv{n_l}{Z_t}} \delta \vec{Z}_t = \vec{\cal L}_t
- {\vec R}_t \sum_{t^\prime\ne t} \pderiv{n_i}{Z_{t^\prime}} \delta
\vec{Z}_{t^\prime}
- {\vec S}_t \sum_{t^\prime\ne t} \pderiv{n_l}{Z_{t^\prime}} \delta
\vec{Z}_{t^\prime},
\end{equation}
which is solved for $\delta(\vec{Z}_t)_d$.

\section{Accelerated lambda iteration}

Although complete linearization is a powerful method, it results in
huge matrices having the dimension $\NF\times\NL$.
On the other hand, the simple $\Lambda$-iteration suffers from
convergence problems.
As an alternative, \cite{fqpt,aqpt} suggested an iterative method based
on the operator splitting method.
This method was later further developed by \cite{dqpt,dqptmrt}.
Basic properties of this method, now called `Accelerated Lambda
Iteration' (ALI) method, were described by \citet*{oab}.
Useful reviews of ALI methods were published by \cite{alirev,alisam}.

We can introduce an
approximate operator $\Lambda^\ast$ representing an approximate solution
of the radiative transfer equation by adding and subtracting the term
$\Lambda^\ast_\nu S_\nu$ at the right hand side of \eqref{lameq},
\begin{equation}
J_\nu = \Lambda_\nu^\ast S_\nu + \zav{\Lambda_\nu - \Lambda_\nu^\ast}
S_\nu 
\end{equation}
In analogy with \eqref{lamit}, we may introduce an iteration scheme
\begin{equation}\label{lamastit}
J_\nu^{(n)} = \Lambda_\nu^\ast S_\nu^{(n)} +
\zav{\Lambda_\nu - \Lambda_\nu^\ast} S_\nu^{(n-1)}
= \Lambda_\nu^\ast S_\nu^{(n)} + \Delta J_\nu^{(n-1)}.
\end{equation}
The basic difference with respect to the $\Lambda$-iteration is that the
$\Lambda^\ast$-operator is now acting on $S_\nu^{(n)}$ and not on
$S_\nu^{(n-1)}$ like in the case of ordinary $\Lambda$-iteration.
The accurate solution of the radiative transfer equation $\Lambda_\nu  
S_\nu$ (the formal solution) is taken from the preceding, i.e.
$(n-1)$ iteration step.
In the limit of the converged solution the equation \eqref{lamastit} is
exact.
The term
$\Delta J_\nu^{(n-1)} = \zav{\Lambda_\nu - \Lambda_\nu^\ast} S_\nu^{(n-1)}$
is the correction term, which describes, how the approximate solution
(obtained with $\Lambda_\nu^\ast$)
for the \emph{current} estimate of the source function $S^{(n-1)}$ differs
from the exact one (obtained with $\Lambda_\nu$).

The normal procedure in multilevel radiative transfer calculations with
ALI is to first determine the correct source function by common solution
of the equations of statistical equilibrium with a simplified expression
for $J_\nu^{(n)}$ \eqref{lamastit}, and then to obtain the radiation
intensity from the calculated source function using the formal solution.
Since the source function is the ratio of emissivity and opacity, which
both depend on population numbers, determination of the source function
means in this case calculation of the population numbers of individual
levels using the equations of statistical equilibrium.
The expression \eqref{lamastit} for the new estimate of the mean
intensity is inserted to the
expressions for the radiative rates
\eqref{radrates},
\begin{subequations}
\begin{align}
n_i R_{il} & = n_i 4\pi \int \frac{\alpha_{il}(\nu)}{h\nu}
\hzav{\Lambda_\nu^\ast S_\nu^{(n)} + \Delta J_\nu^{(n-1)}}
\dnu
\\
n_l R_{li} &
= n_l \frac{g_i}{g_l} 4\pi \int \frac{\alpha_{il}(\nu)}{h\nu}
\zav{\frac{2h\nu^3}{c^2} +
\hzav{\Lambda_\nu^\ast S_\nu^{(n)} + \Delta J_\nu^{(n-1)}}
}
\dnu
\end{align}
\end{subequations}
which are then used in the equations of statistical equilibrium
\eqref{jk2lamitese}.
These equations may be written as
\begin{equation}
\sum_{l\ne i} \szav{n_l \hzav{R_{li} \zav{n_i,n_l} + C_{li}}
- n_i \hzav{R_{il} \zav{n_i,n_l} + C_{il}}} = 0,
\end{equation}
and they are solved for the population numbers $n_i$.
Finally, the formerly linear set
of the equations of statistical equilibrium became a nonlinear one.
This nonlinearity is a tax for simplifying the problem.

There are two basic possibilities how to solve this system of nonlinear
equations, namely the Newton-Raphson scheme (linearization), which
was used, e.g., by \cite{M1}, or more popular approach of
pre-conditioning \citep{mali1}.
The linearization scheme is very similar to that described in the
Section \ref{linearizace}.
However, the savings achieved are very important, since we do not solve
explicitly the problem in frequencies, number of which may reach several
tens of thousands or more.

\subsection{Construction of the $\Lambda^\ast$ operator}

The crucial point is the construction of the operator $\Lambda^\ast$.
It should have two basic properties:
First, it has to be simple enough to be calculated quickly, and, second,
it should describe the most important interactions of the radiation
field with matter as accurately as possible.
\cite{oab} showed that the best choice of such operator is simply the
diagonal of the $\Lambda$ operator.
However, also other construction methods of the approximate operator are
being used.
The tridiagonal operator of \cite{okshort} was used by \cite{klaus3} and
also by \cite{M1}.

The approximate lambda operator has to be numerically consistent with
the formal solution of the radiative transfer equation, so the
\cite{okshort} operator is suitable if we use the short characteristic
solution of the radiative transfer equation.
On the other hand, for Feautrier solution the approximate operator
constructed after \cite{mali1} has to be used.

\section{Summary}

In this introductory part we emphasized the importance of frequency
coupling of radiation (in addition to the spatial and angle coupling)
for the conditions of stellar atmospheres.
This frequency coupling is ensured mainly by equations of statistical
equilibrium, which connect distant parts of the frequency spectrum.
We also briefly summarized underlying equations for the NLTE radiative
transfer problem in stellar atmospheres with a focus on the problem of
its solution for the case of trace elements.

There are two basic methods of the solution of the RTE+ESE system,
namely the complete linearization method, which is used by the Kiel code
\citep[see][\citetalias{Kielcode}]{Kielcode}, and the accelerated lambda
iteration method used by the DETAIL code
\citep[see][\citetalias{DETAILcode}]{DETAILcode}.

\begin{acknowledgements}
The author thanks Drs.~Ad\'ela Kawka and Barry Smalley for their
comments.
This work was partly supported by a GA\,\v{C}R grant 205/07/0031.
The Astronomical Institute Ond\v{r}ejov is supported by a project
AV0Z10030501.
\end{acknowledgements}

\newcommand{\MRT}[1]{in Methods in Radiative Transfer, W. Kalkofen ed.,
        Cambridge Univ. Press., p.~#1}

\newcommand{\AETS}[1]{in The Atmospheres of Early-Type Stars,
        U. Heber \& C. S. Jeffery eds., Lecture Notes in Physics,
        Vol. 401, Springer-Verlag Berlin, p.~#1}

\newcommand{\SAM}[1]{in Stellar Atmosphere Modelling, I. Hubeny,
        D. Mihalas \& K. Werner eds., ASP Conf. Ser. Vol. 288, p.~#1}

\newcommand{\nicesbornik}[1]{in Non-LTE Line Formation for Trace
	Elements in Stellar Atmospheres, R. Monier et al. eds., EAS
	Publ. Ser. Vol. x, p.~#1}
\newcommand{\rokvydani}{2010}

\end{document}